# INFLUENCE OF THE STELLAR MASS FUNCTION ON THE EVAPORATION RATE OF TIDALLY LIMITED POSTCOLLAPSE GLOBULAR CLUSTERS

Hyung Mok Lee

Department of Earth Sciences, Pusan University, Pusan 609-735, Korea

Jeremy Goodman

Princeton University Observatory, Peyton Hall, Princeton, NJ 08544, USA

**ABSTRACT**

We study the rate of escape of stars ("evaporation") from tidally-limited postcollapse globular clusters having a power-law distribution of stellar masses. We use a multi-mass Fokker-Planck code and assume a steady tidal field. Stellar-dynamical processes cause the inner parts of the cluster to expand, which in turn causes stars to overflow the tidal boundary. Mass loss by stellar evolution is assumed to be unimportant in these later evolutionary stages. The fraction of the cluster mass lost per half-mass relaxation time ($t_{\rm rh}$) is roughly constant, in agreement with simple homologous models with equal-mass stars. If $t_{\rm rh}$ is computed in the conventional way from the mean stellar mass, however, a broad stellar mass function can double the loss of mass per $t_{\rm rh}$. We discuss implications of our results for the evolution of globular-cluster systems in our own and other galaxies. In particular, the number of Galactic clusters destroyed by evaporation alone may be as large or larger than the present cluster population.

*Subject headings:* globular clusters – stellar dynamics

## 1. Introduction

The lifetimes of globular clusters are limited by several processes, of which the most securely calculable is probably dynamical relaxation. Stars gradually escape from the cluster potential as two-body encounters replenish the unbound tail of the local velocity distribution (Ambartsumian 1938; Spitzer 1940). This "evaporation" of stars proceeds even in clusters isolated from external potentials (Hénon 1969, Spitzer & Thuan 1972). If one assumes that stars evaporate at the Ambartsumian-Spitzer rate and leave the (isolated) cluster with negligible energy, then the cluster shrivels to zero mass at constant binding energy in $39t_{\rm rh}$, where $t_{\rm rh}$ is the current half-mass relaxation time (King 1957, 1958). Note that this is a time-invariant statement, since $t_{\rm rh}$ decreases with the cluster mass. Galactic globular clusters are tidally limited, however, (*e.g.* Innanen, Harris, & Webbink 1983), except perhaps at large Galactocentric radii (Caputo & Castellani 1984). Therefore it is the mean density of the cluster rather than its binding energy that remains constant as the mass decreases.

The phenomenon of core collapse and its aftermath also influence the evaporation rate. Two-body relaxation removes orbital energy from stars near the core because the velocity distribution is normally "hottest" near the center. This can lead to a runaway increase in the central density—core collapse—as reviewed by Spitzer (1987). Core collapse can be postponed by rapid mass loss following the evolution of the higher-mass stars (cf. von Hoerner 1958; Applegate 1986; Chernoff & Weinberg 1990; henceforth CW) or by primordial binaries (cf. Spitzer & Mathieu 1980, Gao et. al. 1991, Hut et. al. 1992 and references therein). If the cluster is not disrupted by these influences, however, it will eventually suffer core collapse. As the central density rises, binary formation and other dynamical processes beyond simple two-body relaxation become increasingly important. These processes can energize the velocity distribution, and at sufficiently high density they offset the loss of energy from the core by relaxation and prevent further collapse. Since two-body relaxation does not generate orbital energy but merely redistributes it, energy production in the core increases the total energy of the cluster and therefore decrease its binding energy. (The energy we speak of is in the form of stellar motions within the cluster potential. This energy is "produced" at the expense of other forms, such as the binding energy of close binaries or, in some scenarios, nuclear reactions within the stars themselves.) The tidal field of the galaxy enforces a fixed mean density upon the cluster (averaging over the cluster's orbit). As its binding energy decreases, the cluster must shed stars to maintain its density.

Evaporation by the mechanism we have just described is peculiar to postcollapse because the energy-production mechanisms in the core are negligible before core collapse. Additional mass may be lost by processes unrelated to core collapse, such as normal stellar evolution, tidal shocks, close two-body encounters, and diffusion across the tidal boundary that is not driven by energy changes of the entire cluster. We are not concerned here with these other processes. From the theorist's standpoint, the convenience and importance of the postcollapse energy-driven mass loss is that it depends very little on the nature of the energy-producing mechanism(s) in the core. The reason for this is that the core is regulated by the rate at which two-body relaxation can redistribute the energy through the cluster. If energy were produced in the core faster than two-body relaxation could transport it outwards, then the core would expand, and the reduction in central density would eventually shut down energy generation. Therefore the rate of change of the energy of the cluster is ultimately controlled by the relaxation rate near the half-mass radius. [This is true in the time average despite the phenomenon of gravothermal core oscillations (Sugimoto & Bettwieser 1983; Breeden, Cohn, & Hut 1994 and references therein).] As has often been remarked, a postcollapse cluster is in this respect analogous to a main-sequence star, whose luminosity is ultimately determined by opacities rather than nuclear reaction rates.

Comparison of Hénon's (1961) early results with later work confirms the picture described above (cf. also the review by Goodman 1993). Like recent investigators, Hénon used (indeed invented) the orbit-averaged, isotropized Fokker-Planck equation, and he imposed a steady tidal field. He simplified his problem by adopting equal-mass stars and demanding strictly self-similar evolution. The requirement of self-similarity led to vanishing core radius and infinite central density. A central energy flux occurs in his solution as a sort of eigenvalue of the self-similar equations. The nature and even the density dependence of the energy-production mechanism was not specified, though Hénon speculated that binaries might be involved. Hénon's model loses mass at a constant rate, and the lifetime remaining to the cluster is always $22.4 t_{\rm rh}$, where $t_{\rm rh}$ is the half-mass relaxation time defined by equation (3-2). More recently, Lee & Ostriker (1987, henceforth LO) have made fully time-dependent calculations with a specific form for the central source, with a nonsingular (but very dense) core, and with an improved treatment of the tidal boundary conditions. Outside the core, their models evolve much like Hénon's. In particular, well after core collapse their models lose mass at an almost constant rate and have a "life expectancy" $\approx 20 t_{\rm rh}$. We note that LO also assumed equal-mass stars. Thus, although core collapse and tidal limitation have important effects, King's (1958) conclusion is qualitatively confirmed: a (postcollapse) cluster's "life expectancy" against evaporation is an approximately constant multiple of the current $t_{\rm rh}$.

In the past two decades, most theoretical studies have assumed isolated, rather than tidal, boundary conditions. Interest has focused on the evolution of the core. Since the observational discovery that a substantial minority of Galactic clusters appear to have undergone core collapse (Djorgovski & King 1986), however, a few groups have attempted to fit observations of individual clusters to realistic postcollapse models (Lee, Fahlman, & Richer 1991, henceforth LFR; Grabhorn, et. al. 1992; Drukier, Fahlman, & Richer 1992). Fitting photometry and star counts has almost always required tidal limits in order to explain the steepening of the counts in the outskirts of clusters, so these groups have included tidal boundaries in their evolutionary Fokker-Planck models. They have also adopted several stellar mass components. These may be required by the radial light profiles, but they are surely required by deep CCD star counts, which reveal a broad stellar mass function that varies with radius in the cluster (Richer et. al. 1990).

Multicomponent postcollapse models with tidal limits evaporate at approximately constant rates, just as the single-component models of Hénon (1961) and LO do. Expressed as a multiple of $M_{\rm cluster}/t_{\rm rh}$, however, LFR's evaporation rate is three or more times larger than LO's. Since the two groups used very similar assumptions and historically related codes, it seems that multicomponent models evolve more rapidly than single-component ones. The purpose of the present work is to verify this conclusion.

Many investigations have shown that mass mixtures accelerate the evolution of precollapse clusters. The main effect is *mass segregation* (or *stratification*), whereby the heavier populations are selectively enhanced in the core, hastening its collapse (Spitzer 1969, Spitzer & Hart 1971b, Inagaki & Saslaw 1985). There is also an increase in the precollapse evaporation rate with respect to single-component clusters (Hénon 1969, Spitzer & Shull 1975, Johnstone 1993). But as we have noted, the dominant mass-loss mechanism in postcollapse is somewhat different from those that operate before collapse, because of energy-generation in the core. Although multicomponent models have been made of postcollapse gravothermal core oscillations (Murphy, Cohn, & Hut 1990; Grabhorn et. al. 1992), there has not been a systematic study of the effect of the stellar mass spectrum on the evaporation rate and cluster lifetime in postcollapse. That is the goal of this paper. We do not attempt a comprehensive account of star-cluster evolution. However, we will discuss some implications of the enhanced postcollapse evaporation rate for systems of globular clusters.

It has been claimed that the luminosity function of globular clusters is universal among external galaxies (Harris 1991, and references therein). If this is true, then globular clusters can be used as standard candles for distance measurements. Unfortunately, no theoretical justification has been given for the universality of the luminosity function. In view of our poor fundamental understanding of star formation at present, it is certainly possible that the properties of primordial clusters may be almost independent of their host galaxy. Present-day luminosity functions, however, have surely been influenced by selective destruction of the more vulnerable clusters. One expects destruction rates to depend upon the galactic environment (Aguilar, Hut, & Ostriker 1988, henceforth AHO). In our local part of the Galaxy, evaporation is an important destruction mechanism, and it may be the dominant one (AHO).

Hénon (1961) proposed that the present cluster luminosity function might be explained if all clusters were in a dynamical state similar to his selfsimilar model, and he suggested that singular cores could have been overlooked because of the discreteness of cluster stars. It is now recognized that many clusters that had been thought to have nonsingular cores are probably in a postcollapse state (Djorgovski & King 1986, Chernoff & Djorgovski 1989). These appear to be a minority among clusters, though a substantial one. Evaporation is surely not the sole determinant of the cluster luminosity function, but it may well be an important factor. We discuss this point further in §5.

Harris (1991) also addresses the old question whether the entire Galactic spheroid is composed of disrupted clusters. He argues against this idea on two grounds: cluster orbits appear to be more isotropically distributed than those of spheroid stars; and clusters are more metal poor than spheroid stars at the same galactocentric radii ($R$). These are strong arguments. Theory indicates, however, that destruction rates should correlate more strongly with cluster perigalacticon ($R_{\rm p}$) than with present $R$ (e.g., AHO). Therefore, the orbits of surviving globular clusters are expected to be less eccentric on average than those of primordial clusters. If the field stars in the spheroid came from disrupted clusters, their orbits would be more eccentric on average than those of the undisrupted clusters. Thus, the sign at least of the kinematic differences between Population II field stars and clusters is consistent with theoretical expectation.

According to Zinn (1990), the metallicity offset between stars and clusters is small or nonexistent within the Galaxy at $R > 7$ kpc. If cluster metallicities correlate more strongly with $R_{\rm p}$ than with $R$ (Seitzer & Freeman 1981, Seitzer 1983), then clusters would be expected to be more metal-poor than halo stars at the same $R$ because of preferential destruction of the more eccentric, and therefore also more metal-rich, clusters. Under these assumptions, one might also expect the metallicities of local halo *field* stars to increase with decreasing perigalacticon: indeed, Ryan and Norris (1991) have found evidence for such a correlation in the data of Carney et. al. (1990).

While these important issues will be settled only by further observational tests, much could be done by theoreticians to refine their estimates of cluster destruction rates. This paper focuses on one destruction mechanism, postcollapse evaporation, and its sensitivity to the stellar mass function. We describe our numerical methods, initial conditions, and principal assumptions in §2. Section §3 discusses the definition of the half-mass relaxation time, which is used to normalize the evaporation rate, in the presence of unequal stellar masses. We describe our evolutionary models in §4 and discuss their implications for cluster destruction rates in §5.

## 2. Models and Assumptions

We calculate the dynamical evolution of our cluster models with Fokker-Planck code descended from that of Cohn (1979, 1980). The code used in the present paper has been modified by LFR to allow for a multicomponent stellar mass function, a tidal boundary, and three-body-binary heating.

In the present calculations, the mass function is represented by seven discrete mass groups. This is considerably more realistic than a single group, but an even larger number might have been worthwhile.

Evaporation of stars beyond tidal boundary is treated as in LO. In particular, we take the tidal field to be steady and approximate it as spherically symmetric. The first assumption is justified for a cluster on a circular orbit in the disk. On eccentric or disk-crossing orbits, tidal shocks may be important at all phases in the cluster's history (Ostriker, Spitzer, & Chevalier 1972). In general we may expect tidal shocks to accelerate the cluster's evolution by decreasing its binding energy, and perhaps also by spoiling the near-equilibrium form of its distribution function (Spitzer & Chevalier 1973; Spitzer & Shull 1975; CW). Until recently, it was widely believed that tidal shocks affect only the least bound stars, whose period in the cluster is longer than the shock timescale, but Weinberg (1994) has shown that adiabatic invariance does not protect the energies of all of the strongly-bound, short-period stars. We neglect time-dependence of the tide not because it is unimportant, but because to include it would require additional parameters and would obscure the physical process we most wanted to study. The assumption of spherical symmetry in the tidal field is unphysical but inescapable in the one-dimensional code that we use. Stellar orbits near steady and nonsteady tidal boundaries are complex (cf. Keenan 1981, Oh & Lin 1992). The details of the tidal cutoff should not much affect the mass-loss rate in quasisteady postcollapse evolution, however, because the loss rate is driven by the gradual decrease in cluster binding energy, which in turn is controlled by conditions well inside the tidal radius, as described in §1. This expectation is supported by the experiments of LO, who found that large changes in their parametrization of the tidal boundary yielded negligible changes in postcollapse evaporation rate. Although the structure and extent of the outer parts of the cluster are undoubtedly sensitive to the nature of the tidal cutoff, these are not our present concern.

Heating of the stellar velocity distribution in the central parts of the model is attributed to binaries formed in three-body processes. We have included the heating effect of binaries without explicitly following their formation and evolution. Our results are independent of the choice of central energy source, as long as the source provides the energy flux demanded by two-body relaxation near the half-mass radius.

The initial density profile is a King model with dimensionless central potential $W_0 = 7$ and hence a concentration parameter $c \equiv \log(r_{\rm t}/r_{\rm c}) = 1.53$. This is an arbitrary choice: we have few constraints on the primordial structure and concentration of Galactic globulars. Relaxation being an entropy-increasing process, however, postcollapse evolution is insensitive to initial conditions apart from the cluster orbit, the total cluster mass, and the stellar mass function. As a check, we have performed parallel calculations with $W_0 = 4$ and obtained similar results. We assume, of course, that the initial conditions are such as to permit a postcollapse phase. If the initial concentration parameter is too low, the cluster may disrupt before core collapse or it may not reach core collapse in a Hubble time.

A uniform local mass function of Salpeter type is assumed throughout the cluster:

$$N(m)dm = Cm^{-(1+x)}dm, \quad m_{\rm min} < m < m_{\rm max}. \tag{2-1}$$

Thus there is no mass segregation in our initial conditions. We regard $x = 1.35$ as the standard choice, but we have made calculations for a range of values of the exponent $x$. The values of the cutoffs $m_{\rm min}$ and $m_{\rm max}$ are also very important to the evolution. We have taken $m_{\rm max} = 0.7 M_\odot$, the turnoff mass; remnants of more massive stars are ignored. We choose the lower cutoff $m_{\rm min}$ so that $\mu \equiv m_{\rm max}/m_{\rm min}$ is either 3 or 7.

The effect of stellar evolution on the dynamical evolution of globular clusters is an important subject that has not been considered here. CW have demonstrated that globular clusters can be vulnerable to early tidal disruption because of mass loss by rapidly-evolving high-mass stars. The strength of this effect depends on the slope and extent of the initial mass function. In fact, clusters are much less easily disrupted if one uses $m_{\min}$ smaller than Chernoff & Weinberg's value of 0.4 $M_\odot$ (Drukier 1993). In any case, mass loss by unperturbed stellar evolution is relatively unimportant after core collapse. The evolutionary timescale of the turnoff stars is directly proportional to the cluster age, but in a tidally-limited postcollapse cluster, $t_{\rm rh}$ decreases in proportion to the cluster mass, hence linearly in time. If we accept Faber & Gallagher's (1976) stellar-mass-loss rate of $1.5 \times 10^{-11}(L_B/L_\odot)M_\odot\,{\rm yr}^{-1}$, where $L_B$ is the total B-band luminosity of an old stellar population, then the timescale for mass loss from evolved stars is $\tau_* = 7 \times 10^{10} \Upsilon_\odot\,{\rm yr}$, where $\Upsilon_\odot$ is the mass-to-light ratio of the cluster in solar units. The timescale for postcollapse evaporation is $\tau_{\rm evap} \lesssim 22.4 t_{\rm rh}$, which is $\lesssim 2 \times 10^{10}\,{\rm yr}$ since $t_{\rm rh} \lesssim 10^9\,{\rm yr}$ for Galactic postcollapse clusters (§4).

## 3. Half-mass Relaxation Time

This is conventionally defined following Spitzer & Hart (1971a, henceforth SH):

$$t_{\rm rh} \equiv \frac{0.0600 M^{1/2} r_{\rm h}^{3/2}}{G^{1/2} m \log(0.4N)}, \qquad (3\text{-}2)$$

where $M \equiv$ total cluster mass; $r_{\rm h} \equiv$ half-mass radius (the radius of a sphere containing half the cluster mass); $N \equiv$ total number of stars; and $m$ is a characteristic stellar mass. For a cluster with a broad mass spectrum, such as the ones considered in this paper, the actual value of $t_{\rm rh}$ deduced from this expression will depend importantly on the choice of the characteristic mass $m$. This is usually taken to be either the mean mass,

$$\bar m \equiv \left[\int_{m_{\min}}^{m_{\max}} N(m) m\,dm\right] \Big/ \left[\int_{m_{\min}}^{m_{\max}} N(m)\,dm\right], \qquad (3\text{-}3)$$

or something close to the turnoff mass, which we identify with $m_{\max}$. The turnoff mass has the advantage of being readily observable, modulo uncertainties in cluster age, metallicity, and distance. The mean mass, which may better represent the typical star if the mass function is steep, is difficult to determine because of the faintness of the less massive stars, and because mass segregation requires that the mass function be sampled at many radii within the cluster.

Nevertheless, in this theoretical paper, we calculate $t_{\rm rh}$ using the mean mass (3-3) for $m$. Since lighter stars evaporate more quickly than heavier ones, $\bar m$ decreases as our model clusters evolve.

Because this paper is directed towards the dimensionless evaporation rate (4-8), which is inversely proportional to $t_{\rm rh}$, it is useful to examine more closely the assumptions underlying formula (3-2). SH define a local reference relaxation time,

$$t_{\rm rf} \equiv \frac{v_{\rm mf}^3}{3\kappa \psi n_{\rm f} m_{\rm f}^2}, \qquad (3\text{-}4)$$

in which

$$n_{\rm f} \equiv \sum_k n_k, \quad v_{\rm mf}^2 \equiv \frac{\sum_k n_k m_k v_{{\rm m}k}^2}{n_{\rm f}}, \quad m_{\rm f} \equiv \frac{\sum_k m_k n_k}{n_{\rm f}} = \bar m, \qquad (3\text{-}5)$$

where $n_k$ and $v_{{\rm m}k}$ are the local number density and velocity dispersion of stars with mass $m_k$. The quantity $\kappa = 11.8 G^2 \log(0.4N)$, and

$$\psi \equiv \frac{\sum_k m_k^{5/2} n_k}{m_{\rm f}^{5/2} n_{\rm f}}. \qquad (3\text{-}6)$$

TABLE 1

POWER-LAW MASS FUNCTIONS

| | $\mu = 3$ | | $\mu = 7$ | |
|---|---|---|---|---|
| $x$ | $\bar{m}/M_\odot$ | $\bar{\psi}$ | $\bar{m}/M_\odot$ | $\bar{\psi}$ |
| 0 | .425 | 1.188 | .308 | 1.584 |
| 0.5 | .404 | 1.198 | .265 | 1.694 |
| 1.0 | .384 | 1.204 | .227 | 1.755 |
| 1.35 | .372 | 1.204 | .205 | 1.753 |
| 2.0 | .350 | 1.195 | .175 | 1.659 |
| 2.5 | .336 | 1.183 | .159 | 1.541 |

To obtain (3-2), SH take $\psi = 1$, as if all stars had the same mass; they replace $m_f n_f$ by the mean mass density within $r_h$ ($= 3M/8\pi r_h^3$); and they estimate

$$v_{\rm mf}^2 \approx 0.4 \frac{GM}{r_h}. \qquad (3\text{-}7)$$

To the extent that the tidal contribution to the virial equations is small, (3-7) is equivalent to estimating the binding energy of the cluster by $E_{\rm cluster} \approx 0.2 GM^2/r_h$. Although in principle the numerical coefficient in this last relation depends upon the cluster's density profile, in practice it varies by much less than a factor of two over a broad range of pre- and postcollapse theoretical models.

The dimensionless mass moment $\psi$ is substantially larger than unity in our initial models based upon the mass function (2-1). Table 1 indicates the variation of $\bar{m}$ and of $\psi$ with $x$ for our standard ratio of cutoff masses, $\mu \equiv m_{\max}/m_{\min} = 7$. Evidently, $\psi$ is almost independent of the choice of the mass-function slope $x$, but it increases with $\mu$. We have emphasized this point because, as we discuss in §4, the evaporation rate $\xi_e$ is typically a factor of two larger for our multimass models than for analogous single-mass clusters. Incorporating an appropriately-averaged definition of $\psi$ in the definition of $t_{\rm rh}$ would substantially reduce this factor.

## 4. Results

Because of the relatively high central concentration of the initial conditions we have chosen, our models reach core collapse within one half-mass relaxation time. Because of the diffusive nature of dynamical relaxation, the later evolution of our models is largely independent of the initial density profile. It depends mainly on the tidal field, the total initial mass, and the parameters of the initial mass function (2-1): the slope, $x$, and the ratio of mass cutoffs, $\mu \equiv m_{\max}/m_{\min}$.

Although many aspects of the postcollapse evolution of these multicomponent models might be worth exploring, our interest is in the dimensionless evaporation rate

$$\xi_e \equiv -\frac{t_{\rm rh}}{M}\frac{dM}{dt}. \qquad (4\text{-}8)$$

Fig. 1.— The dimensionless evaporation rate (4-8) plotted against time for various mass-function slopes $x$ and cutoff ratios $\mu \equiv m_{\max}/m_{\min}$: (a) $\mu = 3$; (b) $\mu = 7$.

Strictly self-similar evolution requires $\xi_e =$ const. Hénon's (1961) single-component self-similar model has $\xi_e = 0.0446$. Lee & Ostriker (1987), who did not assume self-similarity, and who treated the tidal boundary condition in greater detail than Hénon did, found nearly the same $\xi_e$ in late postcollapse. The run of $\xi_e$ against time for $\mu = (3, 7)$ and for several values of $x$ is shown in Figure 1.

Except where otherwise noted, $t_{\rm rh}$ and $\xi_e$ have been determined using the mean mass of the stars still bound to the cluster. Thus the mean mass changes with time because lighter stars evaporate more rapidly than heavier ones (see below).

The ripples in the $\xi_e$ curves arise from inaccuracies in the iterative recomputation of the cluster potential (cf. Cohn 1980). We have repeated selected runs with more iterations per step, and this reduced the ripples without changing the running average of $\xi_e$. We believe that the sharp upturn at the end of many of the curves is also caused by numerical difficulties in recomputing the potential. The upturn occurs after the cluster has lost more than 90% of its initial mass.

Although $\xi_e$ is not exactly constant in Figure 1, it is approximately so at late times. Clusters with a relatively small range of stellar masses evaporate only slightly faster than single-component model [Fig. 1(a)]. Broader initial functions, however, significantly increase the evaporation rate. At $\mu = 7$, $\xi_e$ is $2-3$ times larger than its value in single-component clusters [Fig. 1(b)].

Fig. 2.— The fraction of the cluster mass remaining, versus time in units of the estimated time of complete evaporation. (a) $\mu = 3$; (b) $\mu = 7$.

The time axis in Figure 1 has been normalized to the extrapolated time $t_{\rm ev}$ at which the cluster entirely disintegrates. Recall that in tidally-limited self-similar evolution, $dM/dt$ itself is constant, so that the cluster mass declines linearly to zero in finite time (King 1958): If one ignores the slow variation of the logarithm in (3-2), then $t_{\rm rh} \propto M/\sqrt{G\bar{\rho}_{\rm h}}$ at fixed $\bar{m}$, where $\bar{\rho}_{\rm h}$ is the mean mass density within $r_{\rm h}$. Self similarity requires that $\bar{\rho}_{\rm h}$ be proportional to the mean density within the tidal radius, which is constant, so $t_{\rm rh} \propto M$. Since $\xi_e$ is constant, it follows from (4-8) that $dM/dt$ is constant, and the time remaining before disintegration is always a constant multiple of the current half-mass relaxation time: $t_{\rm ev} - t = \xi_e^{-1} t_{\rm rh}(t)$.

Although our models are not self-similar, $dM/dt$ is approximately constant, so we can estimate $t_{\rm ev}$ rather accurately, even though the computation cannot be continued until $M \to 0$ (Fig. 2). Figure 3 shows the time remaining before disintegration in units of the current $t_{\rm rh}$, versus $t/t_{\rm ev}$. In self-similar cluster, these curves would be constants, equal to $\xi_e^{-1}$, but since our models are not self-similar, the data in Figure 3 are not completely equivalent to those in Figure 1. These curves are smoother than the earlier ones because they do not require numerical differentiation of $M(t)$; on the other hand, a small error in the estimate of $t_{\rm ev}$ can cause $(t_{\rm ev} - t)/t_{\rm rh}(t)$ to bend sharply upwards or downwards at the end of the evolution. Examining Figures 1&3, one sees that the clusters with relatively narrow mass function, $\mu = 3$, evaporate $25-50\%$ faster than Hénon's (1961) self-similar model. For the broad mass functions with $\mu = 7$, however, the "life-expectancy" of a postcollapse model is $\approx 5 - 8 t_{\rm rh}$ instead of $22.4 t_{\rm rh}$, unless the slope of the mass function is very flat, $x < 1$.

Fig. 3.— The "life-expectancy" of multicomponent clusters in units of instantaneous half-mass relaxation time, versus time. Solid line: 99% (by number) "main-sequence" stars of mass 0.7 $M_\odot$, and 1% "neutron stars" of twice the mass. Dotted line: $x = 0$. Short-dashed line: $x = 1$. Long-dashed line: $x = 1.35$ (Salpeter slope). Dot-dashed line: $x = 2$. (a) $\mu = 7$ (except for MS+NS model). (b) $\mu = 3$.

As expected, lighter stars escape the cluster more rapidly than heavier ones. This is illustrated in the particular case $(x, \mu) = (1.35, 7)$ by Figure 4, which shows the evaporation rate for each mass component separately. The total evaporation rate $\xi_e$ is an average of these individual rates, weighted by the contribution of each component to the mass of the cluster. At early times, $\xi_e$ traces the loss rate of the predominant lighter components; but at late times when the balance of the population has shifted towards the heavier stars, $\xi_e$ reflects their losses.

Since the present-day mass functions of globular clusters are not easily determined observationally, a submultiple of the turnoff mass is often used instead of the unknown mean mass in estimating the half-mass relaxation time. In our models with $\mu = 7$ and $x \in \{0, 1, 2\}$, if we compute $t_{\rm rh}$ using $m = m_{\rm max}$ instead of $\bar{m}$ in (3-2), then the dimensionless life-expectancy $(t_{\rm ev} - t)/t_{\rm rh}$ depends much more strongly on time than it does in Figure 3 and is (necessarily) always larger. In the latter half of the model's lifetime, however, $(t_{\rm ev} - t)/t_{\rm rh} \sim 15$, which is still somewhat smaller than the equal-mass value of 22.4.

We close this section with some results that may bear on the question *why* multicomponent clusters should evolve more rapidly than single-component ones. We suggested in §3 that part of the explanation may lie in the omission of the dimensionless mass moment $\psi$ from the definition (3-2) of $t_{\rm rh}$. The original definition (3-6) of $\psi$ is appropriate only locally, or when mass segregation is absent. A plausible generalization of this quantity to inhomogeneous clusters is

$$\bar{\psi} \equiv \frac{\sum_k m_k^{5/2} N_k}{\bar{m}^{5/2} N}, \qquad (4\text{-}9)$$

where $N_k$ is the number of stars in mass component $k$. Figure 5 displays the evolution of $\bar{\psi}$ in several of our models with $\mu = 7$. Table 1 predicts that $\bar{\psi} \approx 1.7$ initially, but the computed values are somewhat larger because of the discreteness of our seven mass groups. Except for the flattest mass function, $x = 0$, $\bar{\psi}$ is roughly constant until the final stages of cluster disintegration. If we were to reduce $t_{\rm rh}$ by inserting $\bar{\psi}$ in the denominator of (3-2), just as the local moment $\psi$ occurs in the local reference time (3-4), then $\xi_e$ would be increased by the same factor of $\bar{\psi} \sim 2$, and it would then take much the same value in our self-similar models as it does in Hénon's single-component one. It is not clear, of course, that the $\bar{\psi}$ correction is appropriate for a radially inhomogenous cluster.

Fig. 4.— The evaporation rates of individual mass groups for $x = 1.35$, $\mu = 7$. $\xi_{e,k} \equiv -t_{\rm rh} m_k \dot{N}_k$, where $N_k \equiv$ number of bound stars of mass $m_k$.

The evolution rate of a cluster model depends not only upon the relaxation timescale, but also upon the degree to which the distribution function departs from local "thermal" equilibrium. It has been argued that precollapse multicomponent clusters cannot approach equilibrium very closely, even in their cores, unless the mass function is quite narrow or very steep (Spitzer 1969, 1987; Vishniac 1978; Inagaki 1985). We might expect something similar to be true of postcollapse clusters also. Therefore, Figure 6 displays the run of the dynamical "temperature" $kT \propto m_k \overline{v_k^2}$ with radius for each mass component separately, well into the postcollapse phase of our standard model. At the half-mass radius, the spread in temperature among mass groups is $\gtrsim 3$.

Fig. 5.— The dimensionless moment (4-9) of the current mass function versus time, for $\mu = 7$ and $x$ as marked.

## 5. Discussion

We have found the post-core-collapse evaporation rate of clusters with extended stellar mass functions can be 2-3 times greater than that of clusters with equal-mass stars, if the rate is measured with respect to the half-mass relaxation time computed from the mean stellar mass. Even if $t_{\rm rh}$ is computed from the "turnoff" mass rather than the mean mass, the evaporation rate at late times is still slightly larger than the classical equal-mass result.

As remarked in §1, several destructive processes put globular clusters at risk: mass loss and disruption through evolution of high-mass stars, tidal shocks in the Galactic disk, tidal shocks by the bulge, decay of cluster orbits by dynamical friction, and evaporation. Because the initial mass functions, densities, central concentrations, and galactocentric orbits of primordial clusters are unknown, one cannot reliably estimate how many clusters may already have perished.

AHO have estimated the *present* destruction rate by each of the processes mentioned above, using the observed properties of present-day clusters. These rates are clearly subject to fewer observational uncertainties than the destruction rate of a hypothetical primordial population. AHO conclude that evaporation dominates the present-day destruction rate, which they put at ∼ 4 clusters per Hubble time. Although we agree with AHO's general approach and with their qualitative conclusions, we believe, based upon the results reported here, that evaporation is even more important than AHO have estimated.

Djorgovski (1993) lists several physical parameters for all known Galactic globulars. Among these parameters, he provides $t_{\rm rh}$, calculating it for an assumed mean stellar mass $\bar{m} = 0.33 M_\odot$. Djorgovski identifies 30 clusters (out of 143) as postcollapse candidates. Because of inadequate data, no $t_{\rm rh}$ is given for 6 of the postcollapse clusters. Of the remaining 24, all but 2 have $t_{\rm rh} \leq 1.0{\rm Gyr}$. If the cluster system is 13Gyr old, then these 22 clusters will disappear within the next Hubble time.

Our destruction rate of 22 clusters per Hubble time is a conservative estimate because we have considered only a single destructive process, and because we have made estimates only for a limited sample of clusters with postcollapse morphology and known relaxation times. AHO used a different technique to estimate the total destruction rate for their sample of 83 clusters. For each cluster, they estimated a destruction time associated with each of the destruction processes they considered. Then they summed the reciprocals of these times over their entire sample. If we apply this procedure to our sample of 24 postcollapse clusters and consider only evaporation, the result is 132 clusters per Hubble time (= 13Gyr)—comparable to the entire present-day population—whereas AHO found only 4 clusters per Hubble time!

Our cluster mortality rate is much higher than AHO's for several reasons. AHO assigned to postcollapse only those clusters marked "pcc" by Chernoff & Djorgovski (1989), and not those marked "pcc?". Our philosophy is that clusters centrally concentrated enough to be suspected of postcollapse will soon be in postcollapse even if they are not there yet. In fact, all of Chernoff & Djorgovski's "pcc?" clusters have nominal concentration parameters $c \equiv \log(r_{\rm t}/r_{\rm c}) \geq 1.75$. For comparison, our theoretical models collapsed in $< t_{\rm rh}(0)$ starting from $c = 1.5$. (It should be noted, however, that the early evolution of our models is accelerated by mass segregation. Mass segregation has probably been completed in the high-$c$ Galactic clusters.) Therefore we have included Djorgovski's (1993) "c?" as well as his "c" clusters in our postcollapse sample. Also, AHO used $22.4 t_{\rm rh}$ as the evaporative destruction time for their 11 postcollapse clusters, and $40 t_{\rm rh}$ for the rest, whereas we assume $10 t_{\rm rh}$ on the basis of our multicomponent models. On the other hand, we have neglected destruction mechanisms that AHO allow for. In particular, as AHO have pointed out, disk and bulge shocking were probably more important for the primordial cluster population than they are for the present one because of selective destruction of clusters on highly eccentric orbits.

The following simple test supports the conclusion that our "postcollapse" sample will evaporate in $< H_0^{-1}$. The shortest-lived part of the Galactic cluster population should be in a statistical steady state, since it will have been continually replenished from the reservoir of longer-lived clusters. If the steady-state assumption is true of our postcollapse sample, then these clusters should be uniformly distributed in their evaporation times, $\Delta t_{\rm ev} \equiv t_{\rm ev} - t$. Thus for example, there should be twice as many clusters with $\Delta t_{\rm ev} \leq 2{\rm Gyr}$ as with $\Delta t_{\rm ev} \leq 1{\rm Gyr}$. To the extent that $\Delta t_{\rm ev}$ is a universal multiple of $t_{\rm rh}$, the postcollapse sample should also be uniformly distributed in $t_{\rm rh}$. To test this, Figure 7 displays the cumulative distribution of $t_{\rm rh}$ for our postcollapse sample, and compares this with a uniform distribution having the same median $t_{\rm rh}$, $t_{\rm rh,med} = 10^{8.51}{\rm Gyr}$. The observed sample is somewhat more clustered around the median than it should be if it were truly uniform. Also, the theoretical distribution does not allow for any clusters with $t_{\rm rh} > 2 t_{\rm rh,med}$, so the agreement is poor at the high-$t_{\rm rh}$ end; but at this end the steady-state assumption is least reliable, because $\Delta t_{\rm ev} \to H_0^{-1}$. Notwithstanding these minor disagreements, the fit appears very satisfactory overall.

Fig. 6.— Dynamical temperature $m_k \overline{v_k^2}/3$ of mass component $k$ versus radius in model for $x = 1.35$, $\mu = 7$ at $t/t_{\rm ev} = 0.64$.

In summary, we conclude that a significant fraction—not less than a sixth, and probably much more—of the present-day Galactic cluster system will disappear within the next Hubble time. Therefore, the present cluster population may differ systematically from the primordial one. This need not be inconsistent with a universal extragalactic cluster luminosity function. But it would seem to require either that destruction rates as well as primordial cluster properties are constant among galaxies, or that both of these vary among galaxies but conspire to produce similar results today. In our opinion, even the question whether the entire spheroid is formed from disrupted clusters should be regarded as open, because of possible correlations among cluster mortality, metallicity, and orbital eccentricity.

We thank Luis Aguilar for discussions concerning halo metallicity gradients, J.P. Ostriker for discussions of tidal shocks, and an anonymous referee for useful criticisms. J.G. was supported in part by a grant from the David and Lucille Packard Foundation. HML was supported by the Basic Science Research Institute Program BSRI-93-591.

Fig. 7.— Solid line: cumulative distribution of 24 "postcollapse" clusters with respect to half-mass relaxation time. Dashed line: a uniform distribution extending from $t_{\rm rh} = 0$ to twice the median $t_{\rm rh}$ of the solid curve.